\documentclass[journal ]{new-aiaa}
\usepackage{graphicx}
\usepackage{siunitx}
\usepackage{longtable,tabularx}
 \usepackage{float}

\title{Multiscale Reduced-Order Modeling of a Titanium   Skin Panel Subjected to Thermo-Mechanical Loading}

\author{Xiang Zhang \footnote{Assistant Professor, Department of Mechanical Engineering, corresponding author email: xiang.zhang@uwyo.edu, member AIAA} }
\affil{University of Wyoming, Laramie, Wyoming, 82071, US}
\author{Yang Liu  \footnote{Postdoctoral Fellow, Department of Materials,  email: yang.liu@imperial.ac.uk}}
\affil{Imperial College London, Exhibition Road, London, SW72AZ, UK}
\author{Caglar Oskay\footnote{Professor, Department of Civil Engineering,  corresponding author email: caglar.oskay@vanderbilt.edu, member AIAA}}
\affil{Vanderbilt University, Nashville, Tennessee, 37240, US}

\begin{document}

\maketitle

\newcommand{\hilight}[1]{\colorbox{yellow}{#1}}
 \floatstyle{plain}
 \newfloat{Box}{!h}{lob}
 \newcommand{\boxedtext}[3]{
 	\begin{Box} \caption{\small{#1}}
	\hspace{1.cm}
	\fbox{\begin{minipage}[c]{0.92\linewidth} 
	
	\small{#2}
       
       \end{minipage}}
       
       \label{#3}
       \end{Box}
}


\begin{abstract}
This manuscript presents the formulation, implementation, calibration  and application of a multiscale reduced-order model to simulate a  titanium panel structure subjected to  thermo-mechanical loading associated with high-speed flight.  The formulation is based on the eigenstrain-based reduced-order homogenization model (EHM) and further considers thermal strain as well as temperature dependent material properties and evolution laws. The material microstructure (i.e., at the scale of a polycrystalline representative volume element (RVE)) and underlying microstructural mechanisms are directly incorporated and fully coupled with a structural analysis, concurrently probing the response at the structural scale and the material microscale. The proposed approach was fully calibrated using a series of uniaxial tension tests of Ti-6242S  at a wide range of temperatures and two different strain rates. The calibrated model is  then adopted to study the  response  of   a generic aircraft skin panel subjected to thermo-mechanical loading associated with high-speed flight. The analysis focuses on demonstrating the capability of the model to predict not only the structural scale response, but simultaneously  the microscale response, and further studies  the effects of temperature and texture on the response.
\end{abstract}

\section*{Nomenclature}

{\renewcommand\arraystretch{1.0}
\noindent\begin{longtable*}{@{}l @{\quad=\quad} l@{}}
$\mathbf{x}$, $\mathbf{y}$   & Coordinates of the macro- and microscale domain \\
$\zeta$  & Scale separation parameter \\
${\bar {\mathbf{u}}}$,  ${\bar T}$  &    macroscopic   displacement and temperature\\
${\bar {\pmb{\epsilon}}}$, ${\bar {\pmb{\sigma}}}$ ,  $\bar {\mathbf b}$&    macroscopic strain, stress and body force \\
$\bar \rho$, $\bar {c}$,   $\bar K$ &   Homogenized density, specific heat and thermal conductivity at the macroscale \\
$\pmb{\epsilon}^{(\alpha)}$, $\pmb{\mu}^{(\alpha)}$, $\pmb{\sigma}^{(\alpha)}$ &   strain, inelastic strain and stress  of part $\alpha$ \\
$\dot{\gamma}^{s(\alpha)}$, ${s}^{s(\alpha)}$,  ${\tau}^{s(\alpha)}$,  $Z^{s(\alpha)}$  &   slip rate,   strength,  resolved shear stress  and  Schmid factor of the s-th slip system of part $\alpha$ \\     
$s_0^{s(\alpha)}$, $s^{s(\alpha)}_{\mathrm {for}}$,   $s^{s(\alpha)}_{\mathrm{deb}}$ & initial slip resistance, and slip resistance due to forest dislocation and dislocation debris \\
$\rho_{\mathrm {fwd}}^{s(\alpha)}$, $\rho_{\mathrm {rev}}^{s^{+(\alpha)}}$, $\rho_{\mathrm {rev}}^{s^{-(\alpha)}}$ &    dislocation densities of the s-th slip system of part $\alpha$  \\
$\mathbf{A}$, $\mathbf{P}$, $\pmb{\mathcal{A}}$, $\mathbf{M}$ &   coefficient  tensors \\
$\mathbf{I}$ &   Identity tensor \\

\multicolumn{2}{@{}l}{Superscripts}\\
s & slip system index\\
$\alpha$, $\beta$  & part number\\

\multicolumn{2}{@{}l}{Subscripts}\\
$\mathrm{for}$ & forest dislocation\\
$\mathrm{deb}$ & dislocation debris\\
$\mathrm{tot}$ & total dislocation density\\
$i$, $j$, $k$, $l$, $m$, $n$ &   index \\

\end{longtable*}}

\section{Introduction}
\lettrine{H}{igh-speed} airplane  structures are  exposed to  extreme and transient loads associated with aerodynamic environments. These environments involve oscillatory mechanical loading, as well as temporally and spatially varying temperature fields  that  push the structure into a myriad of limit states (such as yielding, fatigue and  buckling), ultimately resulting in fatigue initiation that quickly grows to jeopardize structural survivability.   

Extensive efforts exist in the literature on investigating the response of aerostructures subjected to high-speed flight environments, for the ultimate goal of  structural damage prognosis to assist the design process. In general, these efforts can be grouped into two categories, aiming to address the two most important aspects of this problem. The first category focuses on characterizing the thermo-mechanical loading associated with high-speed flight in order to predict the response of the structure.  \citet{Beberniss:2016a}  and  \citet{Spottswood:2013a}  conducted wind-tunnel experiments to obtain full-field  experimental data for a simple, champed  flat panel exposed to high-speed flow, and measured the temperature  profile as well as displacement over the panel.  Following this effort,  \citet{PerezR:2016a} applied the measured loads to a panel model to simulate its response. \citet{RileyZ:2016b} simulated the structural response to hypersonic boundary-layer transition, where the pressure loads and heat flux are derived from aerothermodynamics, and loosely coupled to a structural dynamic model.   Further developments along this line include  accounting for the impact of  structure deformation on the loads~\citep{ZacharyR:2014a, BrouwerK:2017a, RileyZ:2017a},  fully-coupled multiphysics modeling~\citep{Culler:2011a, GogulapatiA:2014a, GogulapatiA:2015a},    and proper orthogonal decomposition (POD) based reduced-order modeling (ROM) of the elastic structural model~\citep{CrowellA:2012a, GogulapatiA:2017a}. 

The second category focuses on  accurate modeling of the material evolution and damage accumulation in the high-speed flight environment, where relatively limited studies exist  in the  literature. \citet{AryaV:1991Oa} carried out elastic, elastic-plastic, elastic-plastic-creep, and cyclic loading life analysis of a representative hypersonic aircraft cowl lip under thermo-mechanical loading, and further estimated the fatigue crack initiation life using a low-cycle fatigue criteria.  \citet{XueD:1993a} adopted isotropic linear elastic material model, and investigated the nonlinear flutter and fatigue life of simply supported panels using first-order piston theory aerodynamics and arbitrary prescribed temperature distributions.  More recently,  elastic-thermoviscoplastic model~\citep{SobotkaJ:2013a},  the Chaboche constitutive model that accounts for strain hardening together with the Palmgren–Miner linear damage rule~\citep{MinerA:1945a}, viscoelastic–viscoplastic formulation that accounts for oxygen-assisted embrittlement~\citep{OskayC:2010a},     creep and multiplicatively-decomposed plasticity~\citep{CloughJ:2020a} have been used to evaluate the response of aerostructures. However, damage initiation and propagation in metallic material are  highly localized phenomena, and require direct consideration of the material microstructure (i.e.,  at the  scale  of  the polycrystalline representative  volume element (RVE))  and  all  relevant  microstructural deformation mechanisms due to thermo-mechanical loading.  Hence there is a knowledge  and capability   gap in this aspect for accurate structural scale damage prognosis in the current literature for aerostructures under high-speed flight environment.  To fill this gap, one needs to track microstructural mechanisms for capturing failure initiation phenomena under high-speed flight conditions.

Crystal plasticity finite element (CPFE) modeling has been a widely used numerical tool for capturing  a wealth of deformation mechanisms of metallic materials under different loading conditions. Within the context of CPFE, both thermal expansion/contraction due to temperature change~\citep{Meissonnier:2001a, Ozturk:2016a}, as well as temperature dependent flow rules and hardening rules can be straightforwardly incorporated~\citep{Meissonnier:2001a, HuP:2016a, LiuY:2018a, BenzingJ:2019a}. However, the tremendous computation cost associated with CPFE simulations, and the orders of magnitude of difference in the length scale between a polycrystalline RVE and a structural component, make it computationally prohibitive to conduct a structural scale CPFE simulation even with parallel computing techniques~\citep{BishopJ:2015a}. 

To upscale from the microstructure 
 to the structural scale, several homogenization methods have been proposed. These include the Sachs \citep{Sachs:1928a} and Taylor \citep{Taylor:1938a} models,   the grain cluster method~\citep{Tjahjanto:2010a, Yadegari:2015a},  Viscoplastic self-consistent (VPSC) models~\citep{Segurado:2012a, Knezevic:2013a}, the Fourier transform (FFT) method~\citep{Eisenlohr:2013a},  non-uniform transformation field analysis method~\citep{Michel:2016a}, reduced-order variationalmultiscale enrichment method~\citep{ZhangS:2016a}, self-consistent clustering method~\citep{LiuZ:2016a, YuC:2019a} and data driven models~\citep{LatypovM:2017a, LiuZ:2019a}.  Recently, the eigendeformation-based reduced-order homogenization (EHM) model, originally proposed for fiber-reinforced composites~\citep{OskayC:2007a, Crouch:2010a, SparksP:2016a} has been advanced for the modeling of polycrstal plasticity~\citep{ZhangX:2015a}. EHM is based on the transformation field analysis (TFA)~\citep{Dvorak:1992a} and operates in the context of computational homogenization with a focus on model order reduction of the microscale problem. EHM pre-computes certain microstructure information by solving a series of linear elastic problems defined over the  fully resolved microstructure  (i.e., concentration tensors, interaction tensors) and approximates the microscale problem using a much smaller basis spanned over subdomains (also called parts) of the RVE. 
Using this reduced basis, and  prescribed spatial variation of inelastic response fields over the parts, the microscale problem leads to a set of algebraic equations with part-wise responses as unknowns, instead of node-wise quantities as in CPFE. 
EHM has also recently been extended for a  sparse formulation to further improve its efficiency and scalability~\citep{ZhangX:2017c}, and consider different lattice structures and dislocation density based flow rule and hardening laws~\citep{LiuY:2020a}. 
 
This manuscript presents a thermo-mechanical EHM for computationally efficient multiscale analysis of polycrystalline structures subjected to thermo-mechanical loading. In particular, this manuscript presents the following novel contributions:  (1) thermal strain at the microscale is considered as another source of eigenstrain in addition to the plastic strain from dislocation evolution, through the use of thermal influence functions;  (2) an extensive calibration has been conducted against uniaxial tension of titanium alloy at a wide range of temperatures and two different strain rates, leading to an accurate model that could be used for modeling the structural response under thermal-mechanical loading; (3) localized fatigue initiation analysis in a structural  panel subjected to   thermo-mechanical loading associated with  high-speed flight has been performed, indicating the effects of temperature and  microstructural texture on the response at both micro- and structural scales. The panel that is investigated is made of a high temperature titanium alloy, Ti-6Al-2Sn-4Zr-2Mo-0.1Si (Ti-6242S).

The remainder of this manuscript is organized as follows: Section 2 provides an overview of the  EHM  model for mechanical loading only, along   with its extension to account for thermo-mechanical loading.  Model implementation and   calibration  against a series of uniaxial tensile tests of Ti-6242S at various temperatures and different strain rates  are presented in Section 3, followed by the detailed analysis of an aircraft skin panel    in Section 4. Section 5 concludes the manuscript and discusses  the future work.

\section{EHM  for polycrystalline materials subjected to thermo-mechanical loading}

\begin{figure}
	\centerline{
		\includegraphics[width=0.8\textwidth]{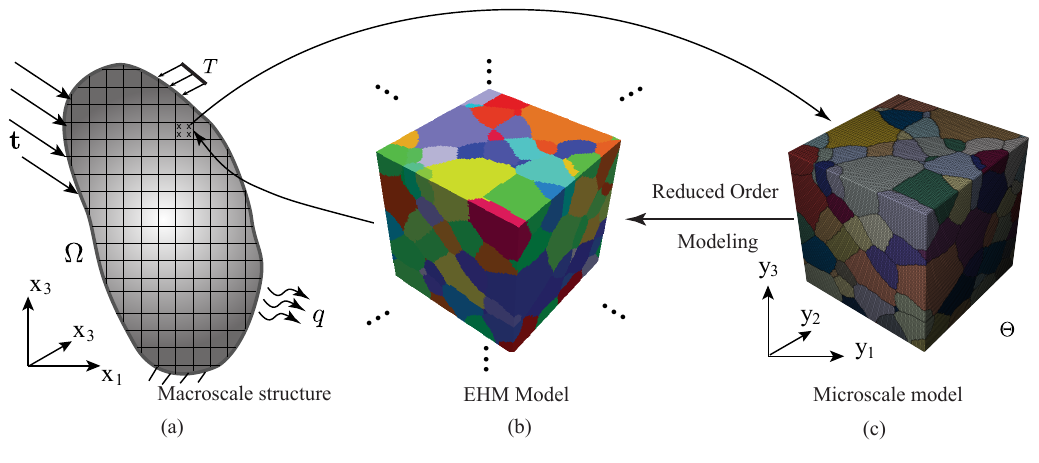}
	}
	\vspace{-0.1in}
	\caption{\small Multiscale reduced-order model.} 
	\label{ProblemSetting} 
\end{figure}
%

\subsection{EHM governing equations}

 EHM is based on the  transformation field theory analysis~\citep{Dvorak:1992a}, to reduce the computational cost associated with the microscale problem in a two-scale computational homogenization (also called FE$^2$, see \citep{Feyel:2000a, Feyel:2003a}) setting as shown in Fig.~\ref{ProblemSetting}.  In  standard computational homogenization, one needs to solve the coupled microscale problem (defined over the microscale domain $\Theta$, characterized by spatial coordinate system $\mathbf{y}$) and the macroscale problem (defined over the macroscale domain $\Omega$, characterized by spatial coordinate system $\mathbf{x}$). EHM aims at  model order reduction of the microscale problem  by: 1) partitioning the microstructure into a small number of subdomains (or parts), where the coarsest partitioning in the context of crystal plasticity is one-part-per-grain; 2) pre-computing certain microstructural information, including concentration and interaction tensors
obtained from linear elastic simulations on the microstructure, associated with the partitioning of the microstructure domain; and 3) prescribing spatial variation of inelastic fields in each part, where uniform response is assumed in each part.   
The macroscale boundary  value problem defining the heterogeneous body  is described by~\citep{OskayC:2007a, ZhangX:2015a}:

\begin{equation}
\bar{\sigma}_{ij, j}({\mathbf x}, T, t) + \bar b_i({\mathbf x}, T, t)=\bar \rho { {\ddot{\bar u}_i}} ({\mathbf x}, T, t) 
\label{MacroScaleEquilibrium}
\end{equation}
and 
\begin{align}
\bar \rho \bar  c \dot {\bar T}(\mathbf{x}, t) +\Big \{\bar{K}_{ij}   [\bar{T}_{,j}({\bf x},  t)]   \Big \}_{,i}=0
\label{O1thermalba}
\end{align}
in which     $\bar {\mathbf{ u}}$ and  $\bar {{ T}}$ denote the macroscopic displacement and temperature fields, respectively. $\bar{\pmb{\sigma}}$ represents the macroscale stress, $\bar{\mathbf{b}}$ the macroscale body force.  $\bar{\rho}$, $\bar{c}$ and $\bar{\mathbf{K}}$ are the homogenized density, specific heat and thermal conductivity,  respectively.    
 While both the displacement and temperature could be microscale dependent (i.e., varying over the microscale domain $\Theta$), we only consider the microscale dependency of the displacement field, neglecting temperature variation over an RVE (i.e., $\bar{T}({\bf x},  t)={T}({\bf x},  t)$). The temperature field  is therefore calculated through a macroscale thermal analysis employing homogenized thermal properties. 
 The macroscale problem is closed with  initial and  boundary conditions, as well as a reduced-order approximation of the microscale problem that effectively provides the macroscale  constitutive responses. The microscale governing equations are expressed in the form of an algebraic system, which is solved for a small set of inelastic strains associated with each part and expressed as (at an arbitrary material point $\mathbf{x}$ on the structure)
\begin{equation}
\dot{\varepsilon}^{(\beta)}_{ij}({\bf x}, T, t)-\sum^n_{\alpha=1}{P^{(\beta \alpha)}_{ijkl}( T)}\dot{\mu}^{(\alpha)}_{kl}({\bf x}, T, t)=A_{ijkl}^{(\beta)}( T)\dot{\bar{\varepsilon}}_{kl}({\bf x}, T, t) +\mathcal{A}^{(\beta)}_{ij}(T)\dot T({\bf x}, t)\label {averesidualderiv}
\end{equation}
in which, $\pmb{\epsilon}^{(\alpha)}$ and $\pmb{\mu}^{(\alpha)}$ represent the part-wise (i.e., subdomain average over $\Theta^{(\alpha)}\subset \Theta$) total strain and inelastic strain tensor of part $\alpha$, respectively.  $n$ is the total number of parts in the RVE.  $\mathbf{P}^{( \beta \alpha)}$ and $\mathbf{A}^{(\beta)}$  are  coefficient tensors expressed as functions of the elastic and inelastic influence functions as described in Ref.~\citep{ZhangX:2015a}, and $\pmb{\mathcal{A}}^{(\beta)}$ is the thermal influence function  introduced following a similar derivation procedure.
 In this manuscript, we use the one-part-per-grain scheme so each part coincides with a grain in the RVE.  The  influence function problems are  summarized in the Appendix for completeness. We note the dependency of the coefficient tensors on temperature, which is different from and is an extension to  the original EHM formulation. This difference is due to the temperature dependent elastic properties, and an efficient evaluation procedure for these temperature dependent coefficient tensors are detailed below. 

\subsection{Viscoplastic constitutive behavior}

 Considering the average inelastic deformation within a grain as a consequence of  crystallographic slip along preferred slip orientations, the inelastic strain rate  in grain (and part) $\alpha$  of the microstructure associated with the macroscale material point $\mathbf x$   is expressed as 
\begin{equation}
\dot{\mu}^{(\alpha)}_{ij}({\bf x}, T, t)=\displaystyle{ \sum_{s=1}^{N} \dot{\gamma}^{s(\alpha)}({\bf x}, T, t)\, Z ^{s(\alpha)}_{ij} }
\label{SheRat}
\end{equation}
where, $\dot{\gamma}^{s(\alpha)}$ is the plastic slip rate on the $s^{th}$ slip system of grain $\alpha$, $N$ is the total number of slip systems, ${\mathbf Z}^{s(\alpha)}$ is the Schmid tensor of the $s^{th}$ slip system of grain $\alpha$,  uniquely describing the orientation of the $s^{th}$ slip system as the dyadic product of the slip direction, $\mathbf{n}^{s(\alpha)}$,  and the normal to the slip plane ${\mathbf m}^{s(\alpha)}$, (i.e.,  $Z^{s(\alpha)}_{ij}=n^{s(\alpha)}_im^{s(\alpha)}_j$). The resolved shear stress acting on a slip system, $\tau^{s(\alpha)}$, is related to  the grain-averaged stress $\pmb{\sigma}^{(\alpha)}$ following the Schmid's law: 
	\begin{equation}
	\tau^{s(\alpha)}({\bf x}, T, t)=\sigma^{(\alpha)}_{ij}({\bf x},T,   t)Z^{s(\alpha)}_{ij}({\bf x})
	\label{SchmidLaw1}
	\end{equation}
	
The slip rate is typically formulated by a flow rule as a function of the resolved shear stress and one or more internal state variables. Here we choose to use a dislocation density based flow rule~\citep{HuP:2016a, LiuY:2018a, LiuY:2020a}
\begin{equation}
\displaystyle{\dot{\gamma}^{s(\alpha)}(\mathbf{x},T, t)=\frac{\rho_m^{s(\alpha)}\nu_{id}^{s(\alpha)} {(b^{s(\alpha)}})^2}{2}\exp\left({\frac{(\tau^{s(\alpha)}-s^{s(\alpha)})\Delta{V}^{s(\alpha)}-\Delta F^{s(\alpha)}}{k T}}\right)\mathrm{sgn}{(\tau^{s(\alpha)})}}
\label{flowrule}
\end{equation}
where, $s^{s(\alpha)}$ is the slip system strength.  $\Delta{V}^{s(\alpha)}$  is the thermal activation volume, $k$ the Boltzman constant. 
$\rho_m^{s(\alpha)}$ is the average mobile dislocation   density, $\nu_{id}^{s(\alpha)}$ the vibration frequency of dislocation segment, $b^{s(\alpha)}$ the magnitude  of the Burgers vector and $\Delta{F}^{s(\alpha)}$ the thermal activation energy. 
The slip system strength  is expressed  as~\cite{Beyerlein:2008a,Knezevic:2013a}
\begin{equation}
s^{s(\alpha)} ({ \mathbf{x},T, t })=s_0^{s(\alpha)}({ \mathbf{x},T }) + s_{\mathrm {for}}^{s(\alpha)}({ \mathbf{x},T, t })+ s_{\mathrm {deb}}^{s(\alpha)}({ \mathbf{x}, T, t }) 
\label{StrengthDecom}
\end{equation}
where $s_0^{s(\alpha)}$  is the initial slip resistance associated with isothermal temperature  introduced by \citet{Ozturk:2016a} based on the work of  \citet{Williams:2002a}.  $s^{s(\alpha)}_{\mathrm {for}}$ and $s^{s(\alpha)}_{\mathrm{deb}}$ are the contributions from the immobile   forest dislocations and dislocation debris, respectively, which constitute obstacles to dislocation mobility. The initial value of  $s_0^{s(\alpha)}$  is calculated with respect to the value at room temperature as
\begin{equation}
s_0^{s(\alpha)}=s_{298K}^{s(\alpha)}-\hat{s}^{s(\alpha)} \left[ 1-\exp \left(\frac{T-T_{\mathrm{ref}}^s}{\widehat{T}^{s(\alpha)}} \right)\right]
\label{S0}
\end{equation}	
where $s_{298K}^{s(\alpha)}$,  $\hat{s}^{s(\alpha)}$, $T^{s(\alpha)}_{\mathrm{ref}}$ and $\hat{T}^{s(\alpha)}$ are material parameters. $s^{s(\alpha)}_{\mathrm{for}}$ and $s^{s(\alpha)}_{\mathrm{deb}}$ 
are expressed as
\begin{equation}
s_{\mathrm {for}}^{s(\alpha)} =\mu^{(\alpha)}\chi b^{s(\alpha)} \sqrt{\rho_{\mathrm{for}}^{s(\alpha)}}
\label{Sfor}
\end{equation}
and
\begin{equation}
s_{\mathrm{deb}}^{s(\alpha)} =\mu^{(\alpha)} b^{s(\alpha)} k_{\mathrm {deb}} \sqrt{\rho_{\mathrm {deb}}^{s(\alpha)}} \ln {\left( \frac {1}{b^{s(\alpha)} \sqrt{\rho_{\mathrm {deb}}^{s(\alpha)}}}\right)}
\label{Sdeb}
\end{equation}
in which, $\mu^{(\alpha)}$ is the shear moduli related to the current temperature, $\chi$ the dislocation  interaction parameter defined as $0.9$ (satisfying the Taylor relationship~\cite{Madec:2002a}). This model does not consider  latent hardening effects, as dislocation density evolution induced by latent hardening has been found to be small compared with that due to self-hardening in the alloy investigated in this study~\citep{OppedalA:2012a,ArdeljanM:2014a,Ardeljan2015}. The latent hardening effects could be added by replacing the scalar $\chi$ with an interaction coefficient matrix as employed in Ref.~\citep{Knezevic:2013b}. $ k_{\mathrm{deb}}$ is a material independent factor associated with low substructure dislocation density and takes the value of 0.086~\citep{Madec:2002a}.  $\rho_{\mathrm{for}}^{s(\alpha)}$ and $\rho_{\mathrm{deb}}^{s(\alpha)}$ are the forest and  dislocation debris densities, respectively.

In the case of  oscillatory  loading, the strain path can potentially change and   the stored forest dislocation can be reversibly reduced by the shearing in the opposite strain path~\citep{Kitayama:2013a,ZecevicM:2015a}. Hence, the total forest dislocation density is physically resolved into three terms as
\begin{equation}
\rho_{\mathrm{for}}^{s(\alpha)}=\rho_{\mathrm {fwd}}^{s(\alpha)}+\rho_{\mathrm {rev}}^{s^{+(\alpha)}}+\rho_{\mathrm {rev}}^{s^{-(\alpha)}}
\end{equation}
where, $\rho_{\mathrm{fwd}}^{s(\alpha)}$ is the forward dislocation density,  and the latter two are the reversible terms corresponding to loading and unloading paths along the $s^{th}$ slip system. 
Experimental observations indicate relatively low initial forest dislocation density in HCP- or BCC-based materials~\citep{Viguier:1995a, Naka:1988a, Follansbee:1989a, ZhangZ:2016a}, and initial forest dislocation density of $\rho_{\mathrm{for},0}=1.0\times10^{12}$ m$^{-2}$ is used. The evolution of the forward dislocation density takes the form
\begin{equation}
\frac{\partial \rho_{\mathrm{fwd}}^{s(\alpha)}}{\partial \gamma^{s(\alpha)}}=(1-p)k_1^{s(\alpha)} \sqrt{\rho_{\mathrm{for}}^{s(\alpha)}}
-k_2^{s(\alpha)} \rho_{\mathrm{for}}^{s(\alpha)}
\end{equation}
where, $p$ is a proportionality parameter determining how much the forest dislocations are reversible, and  chosen based on the  procedure  described in Ref.~\cite{Kitayama:2013a}. Strain is relatively small under the  cyclic loading considered in this study, and more dislocations can be reversed.  $p$  is set as  $0.8$ (following Ref.~\cite{Kitayama:2013a}). $k_1^{s(\alpha)}$  and $k_2^{s(\alpha)}$ are the coefficients controlling the generation of forest dislocations and the annihilation due to dynamic recovery, respectively.

As for the remaining reversible portion of dislocation density, its evolution is different in the slip directions depending on the direction of  shear stress~\citep{Kitayama:2013a}:
\begin{equation}
\frac{\partial \rho_{\mathrm {rev}}^{s^{+(\alpha)}}}{\partial \gamma^{s(\alpha)}}=\begin{cases}
	pk_1^{s(\alpha)} \sqrt{\rho_{\mathrm {for}}^{s(\alpha)}}-k_2^{s(\alpha)} \rho_{\mathrm {rev}}^{s^{+(\alpha)}}  &  \tau^{s(\alpha)} > 0  \\
	-k_1^{s(\alpha)} \sqrt{\rho_{\mathrm {for}}^{s(\alpha)}}{\left( \frac{\rho_{\mathrm {rev}}^{s+(\alpha)}}{\rho_0^{s(\alpha)}}\right)}^{\widehat{m}} &   \tau^{s(\alpha)} < 0
\end{cases}
\end{equation}
and
\begin{equation}
\frac{\partial \rho_{\mathrm {rev}}^{s^{-(\alpha)}}}{\partial \gamma^{s(\alpha)}}=\begin{cases}
-k_1^{s(\alpha)} \sqrt{\rho_{\mathrm{for}}^{s(\alpha)}}{\left( \frac{\rho_{\mathrm {rev}}^{s-(\alpha)}}{\rho_0^{s(\alpha)}}\right)}^{\widehat{m}}  &  \tau^{s(\alpha)} > 0  \\
pk_1^{s(\alpha)} \sqrt{\rho_{\mathrm {for}}^{s(\alpha)}}-k_2^{s(\alpha)}\rho_{\mathrm {rev}}^{s^-(\alpha)} 
 &   \tau^{s,0} < 0
\end{cases}
\end{equation}
where, $\rho_0^s$ is the state of total dislocation density   at the point of the latest load  reversal and $\widehat{m}$ is the dislocation density recombination coefficient taken to be 0.4 for HCC and BCC crystals~\citep{ZecevicM:2015a}. 
If $\tau^{s}=0$, no plastic flow or dislocation density evolution occurs. 

Since the recovery process by means of dislocation climb or cross-slip is related to the  dislocation debris formation, we can describe the evolution of  dislocation debris density to be proportional to the recovery rate. In addition, since hardening at a given slip system is taken to be affected by the total  dislocation debris on all slip systems as suggested by Eq.~\eqref{S0}, the evolution of the dislocation debris  density is expressed as the cumulative contributions from each slip system 
\begin{equation}
d\rho_{\mathrm {deb}}^{(\alpha)}=\sum_s \frac{\partial \rho_{\mathrm {deb}}^{s(\alpha)}}{\partial \gamma^{s(\alpha)}}d\gamma^{s(\alpha)}=\sum_s q b^{s(\alpha)} \sqrt{\rho_{\mathrm {deb}}^{(\alpha)}}k_2^{s(\alpha)} \rho_{\mathrm {for}}^{s(\alpha)}
\label{rhodebevo}
\end{equation}
in which,  $q$ is a recovery rate coefficient, and the initial  dislocation debris density in all slip system is defined to be the same value of $\rho_{\mathrm{deb}, 0}^{s(\alpha)}=1.0\times  10^{10}m^{-2}$~\citep{WilliamsW:1962a,AppelF:1997a}.

\section{Model implementation and calibration}


\subsection{Microstructure reconstruction  and EHM implementation for Ti-6242S}

Ti-6242S is a near-$\alpha$ titanium alloy that has been widely used in aerospace engineering because of its high resistance to failure associated with creep, fatigue and environmental degradation. Its hierarchical microstructure,  such as primary $\alpha$ grains, matrix of lamellar $\alpha$+$\beta$ colonies and microtextured regions of $\alpha$ particles with similar orientations, provides the characteristic features  that contribute to the observed  mechanical and physical properties~\cite{Pilchak:2014a,Pilchak:2016a}. The 2D electron backscatter diffraction (EBSD) scan of the microstructure from Ref.~\citep{Gockel:2016a} is used  to extract the statics to generate a 3D polycrystalline RVE~\citep{LiuY:2020a} using software DREAM.3D~\citep{Groeber:2014a}. A lognormal distribution with $(\sigma, \mu)=(10.21, 0.16)$   \SI{}{\micro\meter} for the $\alpha$ grains and a normal distribution with $(\sigma, \mu)=(0.43, 1.74)$ \SI{}{\micro\meter} for the $\beta$ grains are found to accurately represent the grain area in the experimental EBSD data~\citep{Gockel:2016a}. The grain size statistics, as well as the orientation and misorientation distributions from Ref.~\citep{Gockel:2016a} are used to reconstruct and mesh  the microstructure using the workflow developed in Refs.~\citep{ZhangX:2016a, Phan:2017a}.  This  workflow ensures that  the reconstructed microstructure represents the experimentally observed  material morphology. A 145-grain microstructure shown in Fig.~\ref{Calibration} (a) is used in this study.   

The generated microstructure is then employed to construct the EHM, where each grain constitutes  a EHM part, leading to  145-part  EHM model.  The dislocation density based crystal plasticity model defined in Eqs.~\eqref{SheRat}-\eqref{rhodebevo} together with the EHM framework (Eqs.~\eqref{MacroScaleEquilibrium}-\eqref{averesidualderiv}), and a given set of initial and boundary  conditions fully define the multiscale problem, allowing to simulate a structural scale problem.   The influence function problems are first evaluated on the elastic microstructure using finite element methods following the procedure in Ref.~\citep{ZhangX:2015a}  at different base temperatures (i.e., 295K, 373K, 473K, 589K, 700K, 811K, 873K and 923 K), and   the  corresponding coefficient tensors are computed.  These coefficient tensors at the chosen base temperatures allow us to evaluate coefficient tensors at any intermediate  temperature   following the interpolation procedure in Ref.~\citep{ZhangS:2017a}. 



In this study, the mechanical dissipation induced localized heating due to elasto-visco-plastic processes is considered negligible. This result in a one-way coupling between the thermal and mechanical problems. We   employed a staggered solution scheme to evaluate the coupled thermo-mechanical  problems.  The temperature history is determined first  by solving the steady-state thermal problem. Then the mechanical problem with the determined temperature history is solved with the EHM for the Ti-6242S serving as constitutive law. 
The EHM implementation has been performed for implicit  quasi-static and dynamic analyses.  At a given  time increment of any structural scale integration point, we  track the response of  the   145-grain microstructure, using the temperature at that particular integration point and time increment. This allows us to  probe not only the structural scale response, but also the microscale (i.e., grain scale) response throughout the structural domain.

\subsection{Model calibration for Ti-6242S}
EHM has been  previously verified against direct  CPFE simulations  on different microstructure configurations and subjected to various loading conditions~\citep{LiuY:2020a}. 
This section aims at the calibration of the material parameters using strain controlled uniaxial  tension of Ti-6242S at eight different temperatures (between $295$K and $923$K) and two different strain rates (quasi-static (QS): $8.33\times10^{-5}$/s; high strain rate (HS): 0.01/s)  to arrive at a set of material parameters that provides an overall satisfactory match between simulations and experiments for all the available tests.  More details about the experimental tests can be found in Ref.~\citep{Gockel:2016a}. 
It should be noted that, while the experimental tests are at different temperatures, the load is applied after the specimen rises to the specified temperature and the temperature remains constant during the loading.  This indicates, during these calibrations, the thermal expansion contribution in Eq.~\eqref{averesidualderiv} is negligible.  The thermal expansion is accounted for structural scenarios where temperatures vary in space and time, or when boundary conditions induces thermal stresses  (see Section ~\ref{StructuralAna}).   


~\citet{Ogi:2004a} measured the elastic constants of $\alpha$ titanium at various temperatures and reported a linear dependency of elastic constants on temperature between room temperature and approximately $1000$K.  Based on these findings, \citet{Ozturk:2016a} calculated the slopes corresponding to the reduction of elastic constants with temperature increase for both $\alpha$ and $\beta$ titanium   to account for the temperature effects in CPFE simulation of Ti-6242.    The thermal expansion for $\alpha$ phase is taken to be anisotropic as $\alpha_{<a>}=1.8\times 10^{-5}/$K and  $\alpha_{<
c>}=1.1\times 10^{-5}/$K and for the $\beta$ phase, isotropic expansion is considered with $\alpha =0.9\times 10^{-5}/$K~\citep{Ozturk:2016b}.

 The dislocation mobilities in basal, prismatic and pyramidal slip systems of HCP and three types of BCC crystals are implemented in 
the current model, 
and same parameters for the three types BCC slip systems are assumed similar to that in Ref.~\citep{LiuY:2018a}. This simplification is reasonable considering the  low volume fraction of the $\beta$ phase in the material microstructure. 

\begin{figure}
\centerline{
	\includegraphics[width=0.85\textwidth]{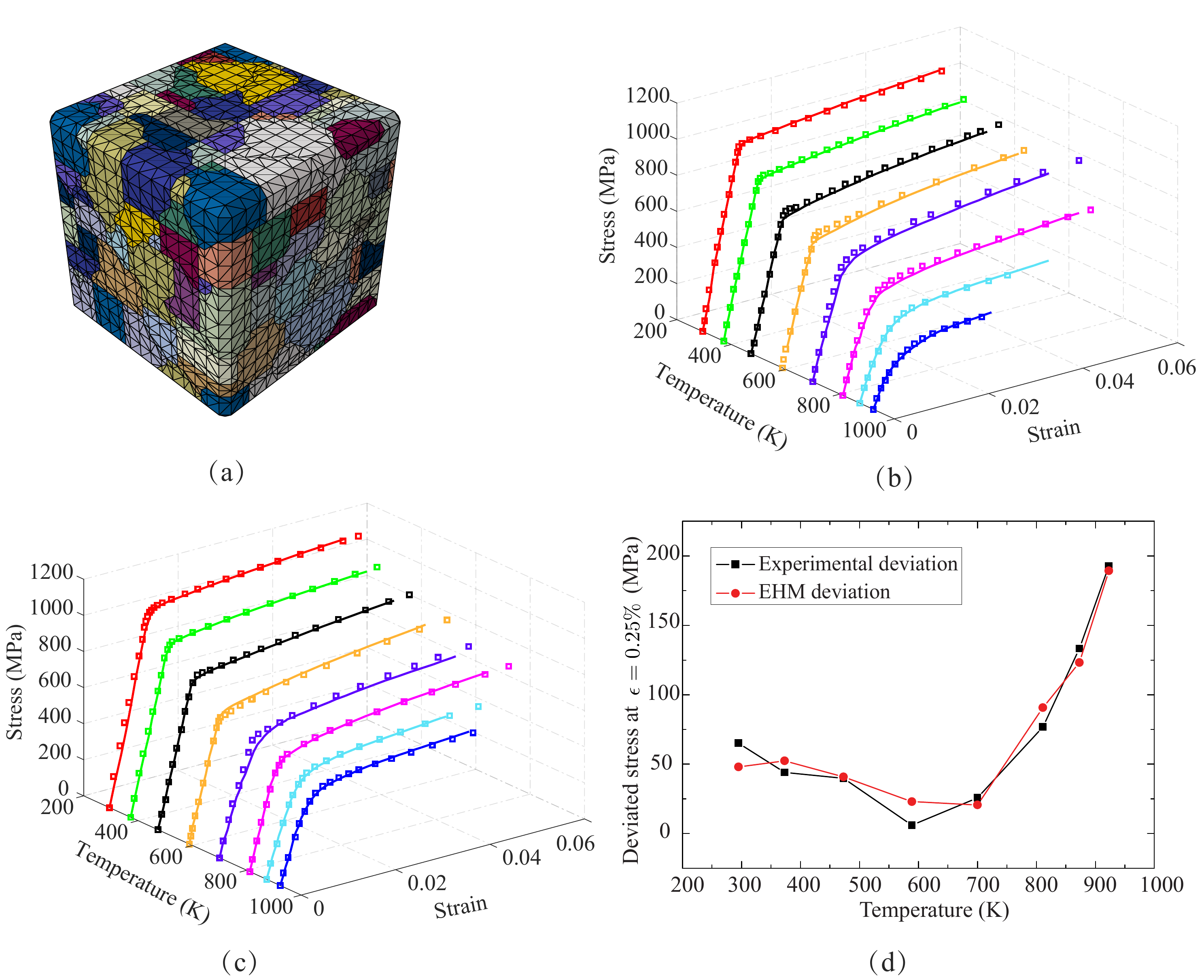}
}
\captionsetup{width=.8\textwidth}
\caption{\small Validation of the EHM model: (a) a 145-grain microstrucre used in this study; (b) simulated QS tests at different temperatures compared with experiments; (c) simulated HS tests at different temperatures compared with experiments; (d) Stress deviation between the two strain rates  as a function of  temperature.} 
\label{Calibration} 
\end{figure}

Due to the large number of parameters used in the current model, a special procedure is adopted to calibrate the  model. First, a physically plausible   range is determined for each parameter based on literature study~\cite{Deka:2006a,Ghosh:2011a,Ghosh:2013a,Ozturk:2016a} and preliminary simulations. 
  Two strain rates and eight temperature conditions ranging from 298 to 923 K are then used to perform calibration separately, leading to a smaller sub-range for each parameter for each single test. Consequently,  temperature- and strain-rate-consistent parameters are obtained in the intersection of the parameter sub-ranges that fits all experiments.  The calibrated model parameters are shown in Table \ref{CpehmPara}.  Using these parameters, the stress-strain   responses under different loading conditions and temperatures are compared to the experiments in Fig.~\ref{Calibration} (b)-(c). The maximum error of overall stress-strain curves is less than 9\%. 
  It is noted that the ultimate strength decreases as temperature increases for both strain rates. 
  To understand the effects of strain rates, the stress deviations between the QS and HS tests at 0.25\% strain are plotted at different temperatures as shown in Fig. \ref{Calibration} (d). It is  shown that the strain rate sensitivity is higher at high and low temperatures, but relatively low at intermediate temperature range of [600K, 700K]. This non-trivial rate sensitivity is well captured by the current model. 

\begin{table}[ht]
\small
\caption {\small Calibrated EHM  parameters for Ti 6242S.}
\centering
\begin{tabular}{ | p{1.15cm}|p{1.1cm}|p{2.15cm}|p{2.25cm}|p{2.35cm}|p{3cm}|p{2.15cm}| } 
 \hline
 \multicolumn{2}{|c|}{Parameters} &  \multicolumn{5}{|c|}{Slip Systems} \\
 \hline
Symbol & Unit & Basal $\langle a \rangle$& Prismatic $\langle a \rangle$ & Pyramidal $\langle a \rangle$ & Pyramidal $\langle c+a \rangle$ & BCC Slip \\
  $\Delta F^\alpha$           & J                           & $3.95 \times 10^{-20}$  & $3.81 \times 10^{-20}$ & $4.27 \times 10^{-20}$ & $4.73 \times 10^{-20}$ & $3.74 \times 10^{-20}$ \\
  $\Delta V^\alpha$           &  m$^3$                    & $5.91 \times 10^{-29}$ & $8.20 \times 10^{-29}$ & $7.40 \times 10^{-29}$ & $8.85 \times 10^{-29}$ & $6.30 \times 10^{-29}$ \\
 $k$                                    &  J$\cdot$ K$^{-1}$  & $1.38 \times 10^{-23} $ & $1.38 \times 10^{-23} $& $1.38 \times 10^{-23}$ & $1.38 \times 10^{-23}$ & $1.38 \times 10^{-23}$\\
  $\rho_m$                         &  m$^{-2}$                  & $5.00\times 10^{12}$  &$5.00\times 10^{12}$ & $5.00\times 10^{12}$ & $5.00\times 10^{12}$ & $5.00\times 10^{12}$\\
 $\nu_{id}$                       &  Hz                     &$1.00\times 10^{12}$  &$1.00\times 10^{12}$& $1.00\times 10^{12}$ &$1.00\times 10^{12}$ & $1.00\times 10^{12}$\\
 $b^\alpha$                          &  $\mu$m               &$2.94\times 10^{-4}$  & $2.95\times 10^{-4}$  & $2.95\times 10^{-4}$  &$4.68\times 10^{-4}$  & $2.86\times 10^{-4}$ \\
 $s^\alpha_{0,ini}$             &  MPa               &500  & 435 & 680 &677 &732\\
 $s^\alpha_{298K}$            &  MPa               &245  & 200 & 180 & 180 & 300\\
 $k^\alpha_1$                       &  m$^{-1}$      &$1.80 \times10^7$  &$1.68 \times10^7$ & $1.67 \times10^7$ &$2.40 \times10^7$ &$ 1.03 \times10^7$\\
 $D^\alpha$                          &  MPa            &300  & 330& 100 &90 & 230\\
 \hline
\end{tabular}
\label{CpehmPara}
\end{table}

\section{Investigation of a   skin panel subjected to thermo-mechanical loading associated with high-speed flight}
\label{StructuralAna}

In this section, the calibrated EHM model for Ti-6242S is utilized to analyze a representative skin  panel structure subjected to combined thermo-mechanical loading associated with a high-speed  flight condition.  This analysis aims at computationally characterizing the deformation processes at both  structural and microscopic scale  that could  be potentially used for   failure initiation analysis and structural damage prognosis. 

\subsection{Representative  skin panel structure}

\begin{figure}
	\centerline{
		\includegraphics[width=0.9\textwidth]{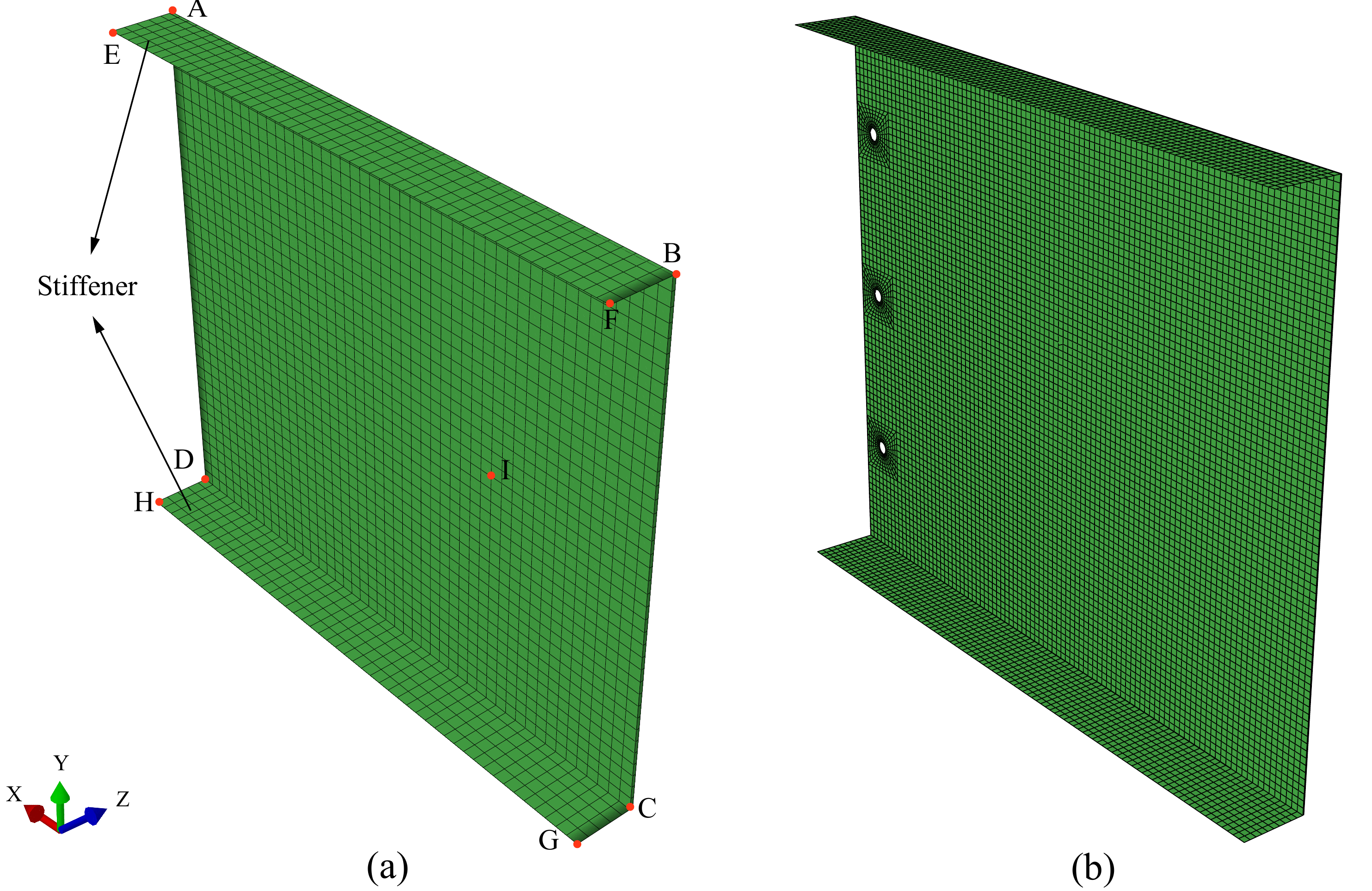}
	}
	\captionsetup{width=.8\textwidth}
	\caption{\small  Geometry and discretization of the two panels: a) Skin panel discretized with the original mesh used in Refs.~\citep{Culler:2010a, Culler:2011a} (element size  $h=6.35$ mm); Bolted panel panel discretized with refined mesh  ($h=3.175$ mm).} 
	\label{PaneGeom} 
\end{figure}

 A generic skin panel  that has been previously  investigated in the literature~\citep{Culler:2010a, Culler:2011a, Spottswood:2013a, GogulapatiA:2014a, GogulapatiA:2015a} is used in this study. 
The geometry and the mesh of the panel is shown in Fig.~\ref{PaneGeom} (a). 
The panel has a dimension of $305.8 \times 254 \times 0.5$ mm, with stiffeners along the long edges,  and is  made of Ti-6242S. The width of the stiffeners is 30 mm and the  thickness is 0.021 mm.  The panel is discretized with  tri-linear eight-noded hexahedron elements with reduced integration that are regularized with hourglass stiffness, and the stiffeners are discretized with  four-noded shell elements  with reduced
integration. A modified version of the panel, with refined mesh and three bolt holes as shown in Fig.~\ref{PaneGeom} (b) is also developed and used in this study as further discussed.   

In the current simulations, the stiffeners are taken to be elastic, while for the panel the EHM for Ti-6242S based on the 145-grain microstructure is adopted.  The simulations using the  mesh in Fig.~\ref{PaneGeom}(a) track the response of  278,400 grains in addition to the structural scale response. The panel is subjected to thermo-mechanical loading associated with a Mach 2, free-stream dynamic pressure of 123 KPa~\citep{Spottswood:2013a}. 



\begin{figure}[h]
	\centerline{
		\includegraphics[width=0.85\textwidth]{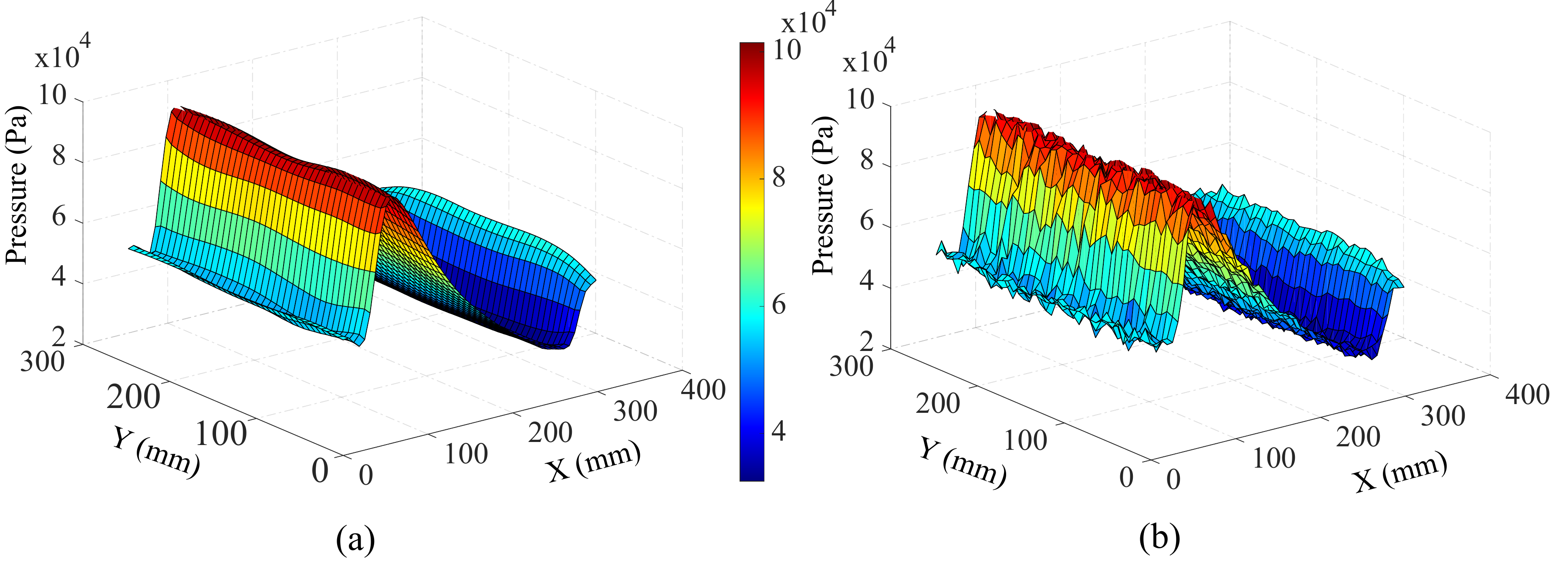}
	}
	\captionsetup{width=.85\textwidth}
	\caption{\small  Pressure profile: (a) mean pressure; (b) total pressure at time 0.01s.} 
	\label{presureprofile} 
\end{figure}

The pressure applied on the panel  consists  of a mean pressure and a temporally and spatially varying acoustic part. The mean pressure is obtained from a steady state  CFD calculation using Reynolds-Averaged Navier-Stokes (RANS) equations for flow on a rigid panel (see Ref.~\citep{Culler:2010a} for details)  as shown in Fig.~\ref{presureprofile} (a).  The acoustic  part is the  turbulent boundary layer (TBL) pressure on the panel, approximated  using a series of random  cosine functions  to generate the spatial and temporal distributions. The number of cosine functions and magnitude of each cosine function are estimated from the work of ~\citet{Spottswood:2013a}, while  frequency are randomly chosen between 1 and 500 Hz.  
{This pressure profile  shown in Fig.~\ref{presureprofile}  is consistent with the reported numerical and experimental results } in Refs.~\citep{GogulapatiA:2014a, GogulapatiA:2015a}.

\begin{figure}
	\centerline{
		\includegraphics[width=0.9\textwidth]{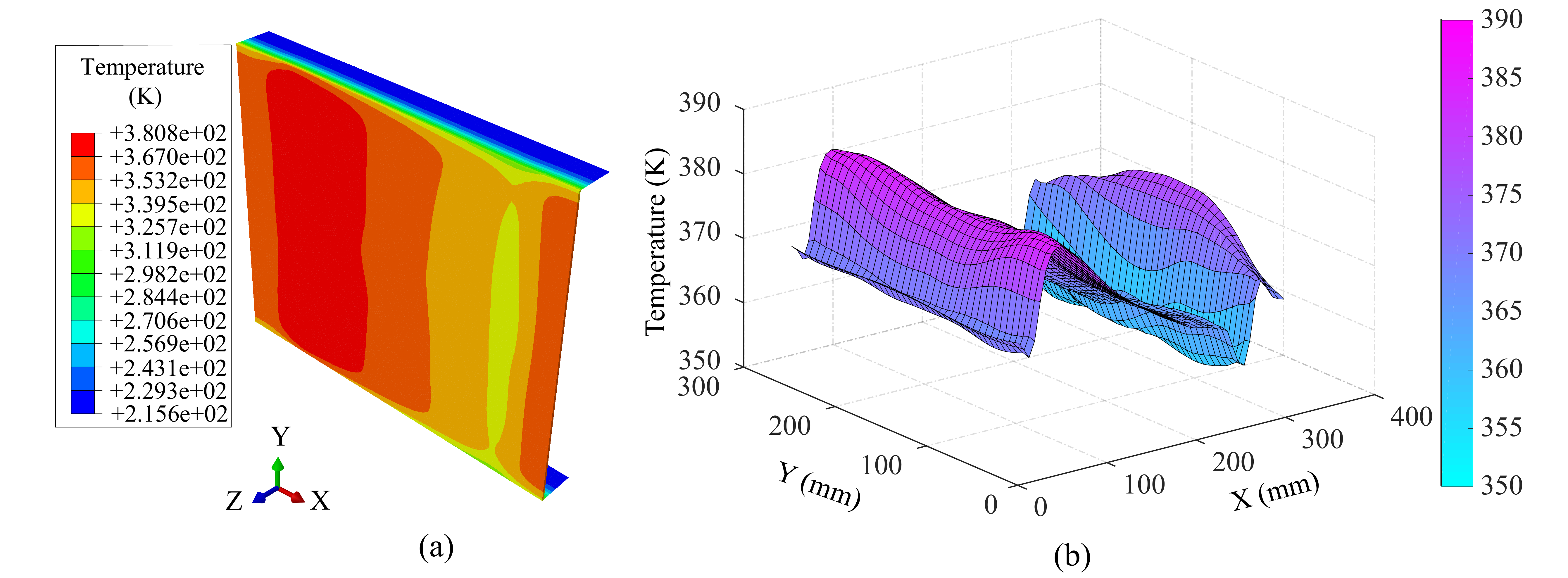}
	}
	\captionsetup{width=.8\textwidth}
	\caption{\small  Temperature profile of the panel at t=20 s: (a) temperature contour from thermal analysis; b) visualization in 3D. } 
	\label{TempProfile} 
\end{figure}

To obtain the spatially varying temperature profile over the panel, a thermal analysis of the panel subjected to a free stream temperature is used. The analysis considers  aerodynamic heating and convective and  radiative heat conduction on the panel surface~\citep{Culler:2010a}. The adiabatic wall temperature  is approximated by performing separate inviscid and viscous flow solutions  following the work of~\citet{AndersonJ:1998a}. In the current study, we start with the temperature profile of the panel at t=20 s after imposing the  adiabatic wall temperature, and conduct a thermo-mechanical analysis for 20 ms. At the beginning of the simulation (i.e., 20 s after aerodynamic heating), the calculated temperature profile for free stream temperature of 388 K is   shown in Fig.~\ref{TempProfile}, with a temperature range of approximately 300-385 K. This temperature profile has a similar pattern with that in Ref.~\citep{Matney:2014a}. We note that since within the current temperature range the change in elastic moduli is negligible,  we directly use the values at room temperature from~\citep{Deka:2006a} to save computational cost. For a different application where the temperature range is significant,  the slopes reported in Ref.~\citep{Ozturk:2016a} corresponding to the reduction of elastic constants with temperature increase for both $\alpha$ and $\beta$ titanium   are  used to consider the temperature dependent elastic constants. Within the   20 ms span of the analysis, temporal temperature change is relatively small. We hence consider  a steady state  but spatially varying temperature field over the structural domain (i.e., $\dot {{T}}=0$ in Eq.~\eqref{averesidualderiv}).

\subsection{Time step size and mesh size convergence study}

\begin{figure}[h]
	\centerline{
		\includegraphics[width=0.45\textwidth]{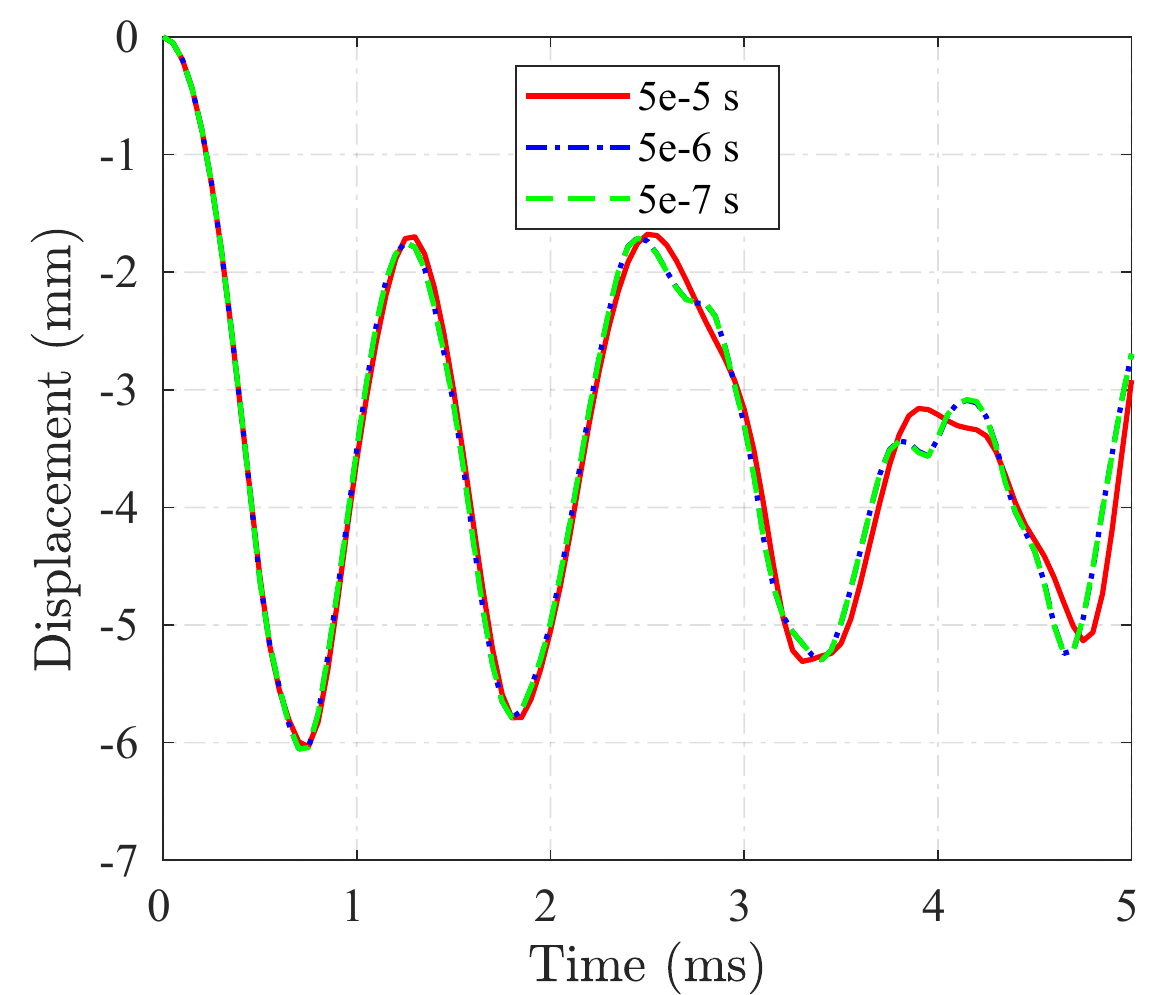}
	}
	\captionsetup{width=.85\textwidth}
	\caption{\small  Displacement history of a chosen node (point I in Fig.~\ref{PaneGeom}(c))  from the linear elastic simulations with different time step sizes.} 
	\label{TimeStepConvergence} 
\end{figure}

Due to the high frequency loading considered in the current study, a time step and mesh size convergence study has been conducted to choose the appropriate element and time step sizes that provide convergent and most efficient simulations. Since simulations are performed using implicit analyses, time step and mesh size selection affect solution accuracy only. Time step stability, a common concern in explicit analysis is not a factor in the implicit approach.    Three different time step sizes, 5e-5 s, 5e-6 s and 5e-7 s are used to  study the response of the panel on the original mesh in Fig.~\ref{PaneGeom} (a). The obtained displacement history  of a  high pressure location (highlighted point I in Fig.~\ref{PaneGeom} (a))  from different time step sizes are shown in Fig.~\ref{TimeStepConvergence}.  The figure indicates that  the  time step size of 5e-6 s yields a converged response, hence the remaining simulations   are based on the  time step size of 5e-6 s. 

\begin{figure}
	\centerline{
		\includegraphics[width=0.9\textwidth]{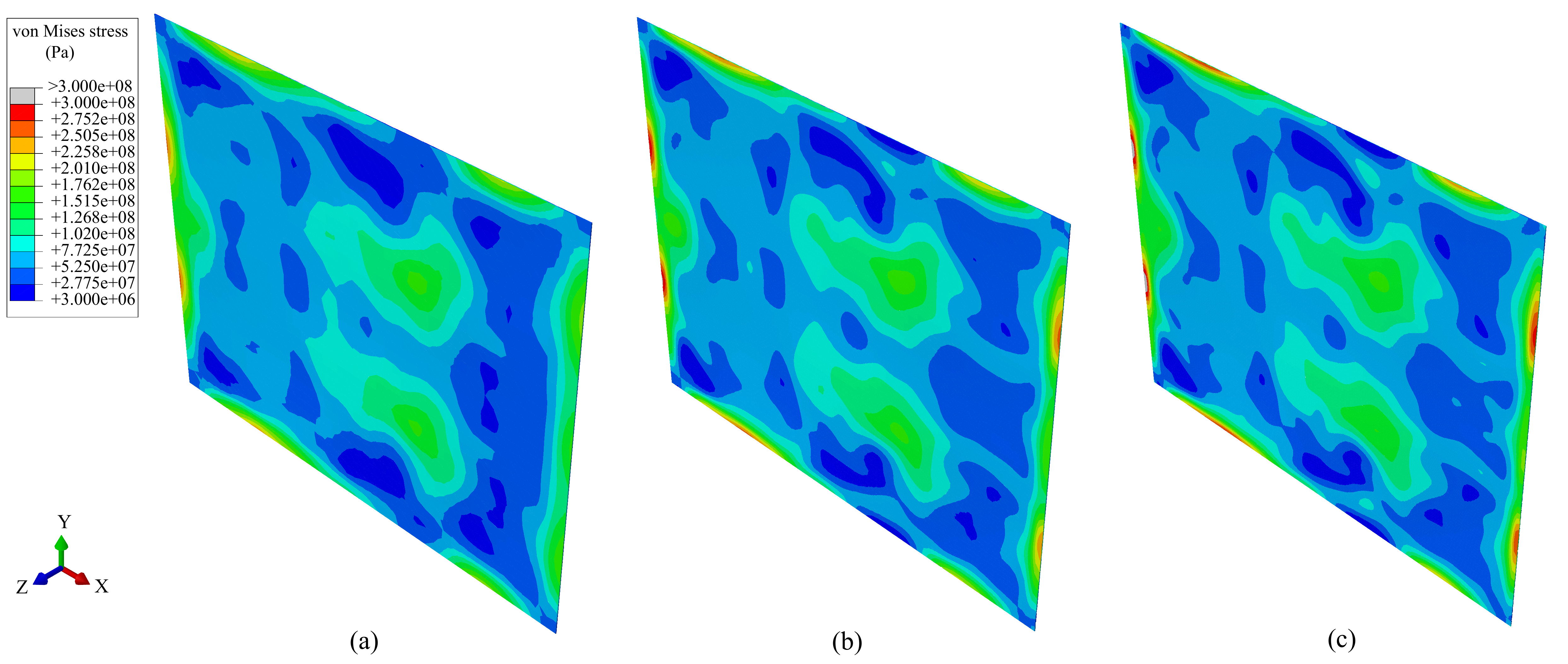}
	}
	\captionsetup{width=.8\textwidth}
	\caption{\small  Stress contours of the panel at $t=1$ ms with different mesh size: (a) $h=6.35$ mm; (b) $h=3.175$ mm; (c) $h=2.17$ mm; } 
	\label{MeshSizeConvergence} 
\end{figure}

We also conduct a mesh size convergence study   by considering two additional  mesh sizes where the  element size is set to  one half ($h=3.175$ mm) and one third ($h=2.11$mm) of that of the original mesh shown in Fig.~\ref{PaneGeom} (a). The stress contour at time = 5 ms from the simulations using different mesh sizes are shown in Fig.~\ref{MeshSizeConvergence}.  The results show relatively small differences, and the intermediate mesh density (i.e., $h=3.175$ mm)  is chosen for all subsequent simulations.  Using this mesh, the simulation tracks the response of 2,818,800 grains throughout the structural domain.

\subsection{Structural scale response investigation: effects of temperature}
 To understand the effect of temperature on the response of the panel (Fig.~\ref{PaneGeom} (a)), we consider different cases with uniform and spatially varying temperatures. In the uniform case,  the temperature of the structure is  473.15 K everywhere. In the spatially varying case, the temperature profile shown in Fig.~\ref{TempProfile} is used. We refer these simulations as Case 1 and  Case 2, respectively. To promote the plastic deformation, a factor of 5 is multiplied to the pressure profile for this panel without bolt holes.

In addition to the macroscale and microscale quantities, we also track additional derived quantities from the microscale response.  \citet{JosephS:2018a,JosephS:2018b}  observed that there are intense dislocation pile-ups at the grain boundaries slightly underneath the crack nucleation surface of fatigued Ti-6242S using transmission electron microscopy. This observation  leads to the proposal of using the dislocation density discrepancy between neighboring grains as a fatigue indicator parameter (FIP) for Ti-6242S under cyclic loading~\citep{LiuY:2020a}.    This  FIP  finds the maximum sessile dislocation density discrepancy across all grain pairs within the microstructure, $\Delta \rho_{\mathrm{tot, max}}$. To identify $\Delta \rho_{\mathrm{tot, max}}$, the largest total sessile dislocation density  across all systems of  grain $i$,  $(\rho_{\mathrm{tot}})^i$,  is computed for all grains as
\begin{equation}
(\rho_{\text{tot}})^i =  \max_{s\in \{ 1, 2, ..., N\}} \left\{ (\rho_{\text{tot}})^{i}  \right\}
\end{equation}
Then  $\Delta \rho_{\mathrm{tot, max}}$ is identified as 

\begin{equation}
\Delta\rho_{\text{tot, max}} = \max\limits_{i\in\{1,...,n\}}\left\{ \max\limits_{j\in\{1,...,m_i\}} \left\{  \left| (\rho_{\text{tot}})^{i} - (\rho_{\text{tot}})^{k(i, j)} \right| \right \} \right\}
\label{MD3}
\end{equation}
in which, $m_i$  is the total number of neighbors of grain $i$, and  $k(i, j)$ is the grain ID for the $j^{\textrm{th}}$ neighbor of grain $i$. 
$\Delta \rho _{\mathrm{tot, max}}$ is the critical state of  grain boundary pile-up for a grain and indicates potential for fracture initiation under oscillatory loading.   We also define the averaged   plastic strain ($\bar \epsilon_{ij}^p$) of the RVE attached to a macroscale integration point as
\begin{equation}
\bar \epsilon_{ij}^p=\sum_{\alpha=1}^n  C^{(\alpha)}\epsilon_{ij}^{p(\alpha)}
\end{equation}
in which, $\pmb{\epsilon}^{p(\alpha)}$ denotes the plastic strain in grain $\alpha$ with $C^{(\alpha)}$ being the volume fraction of grain $\alpha$. These averaged plastic strains are used to calculate the equivalent plastic strain ($\bar \epsilon^{\mathrm{eqp}}$) following~\citep{ChenWF:1988a}

\begin{equation}
\bar \epsilon^{\mathrm{eqp}}=\sqrt{\frac{2}{3}\bar \epsilon^{p}_{ij} \bar \epsilon^{p}_{ij}}
\end{equation} 

\begin{figure}
	\centerline{
		\includegraphics[width=0.9\textwidth]{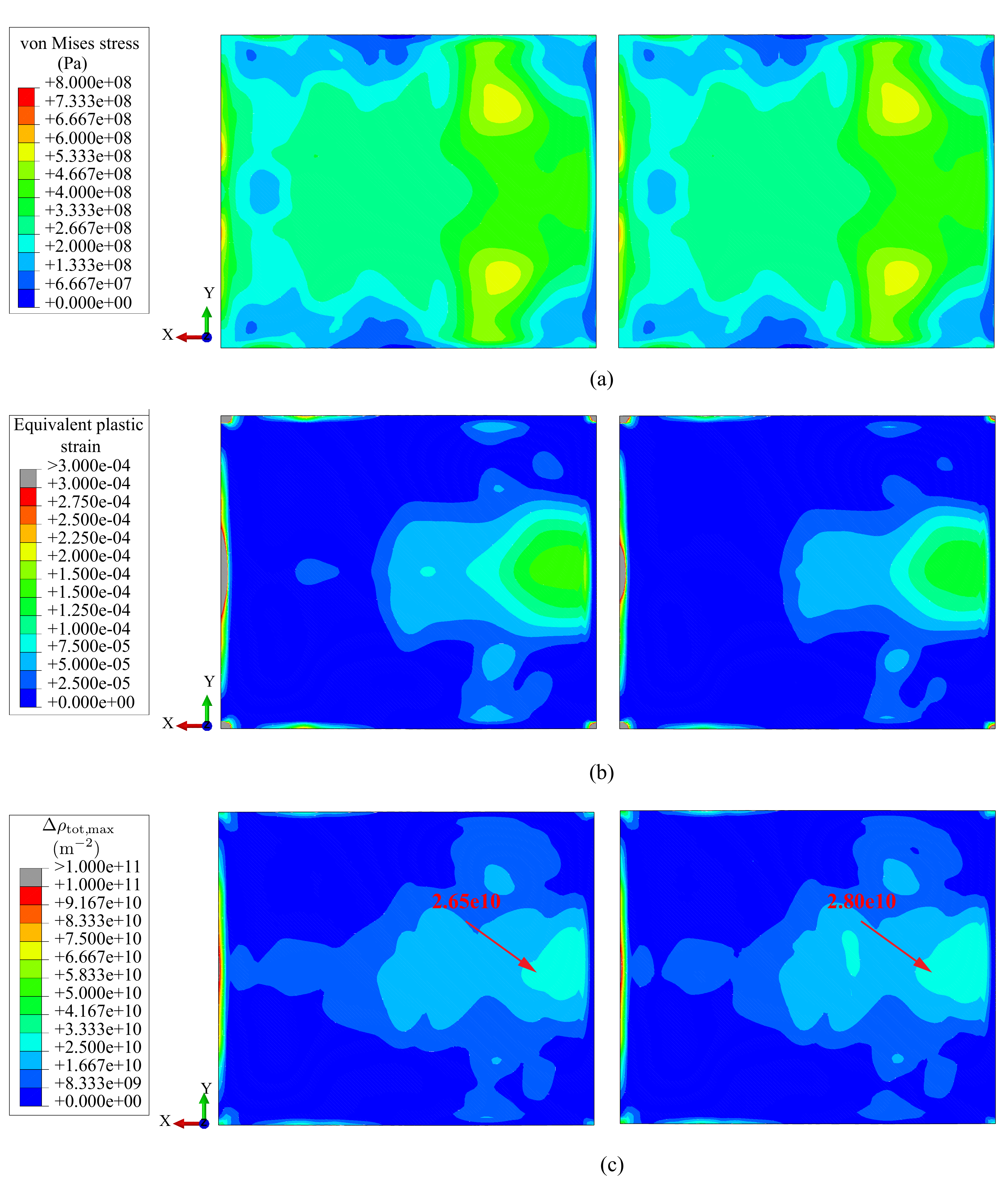}
	}
	\captionsetup{width=.8\textwidth}
	\caption{\small  Comparison between Cases 1 (left) and 2 (right): a) von Mises stress; b) equivalent plastic strain and;  c)  maximum dislocation density discrepancy.} 
	\label{EffectofTemp} 
\end{figure}

 Fig.~\ref{EffectofTemp} provides the  comparison of the von Mises stress, equivalent plastic strain as well as the  maximum dislocation density  discrepancy in Cases 1 and 2. The stress comparison in Fig.~\ref{EffectofTemp} shows that the primary  stress concentration region is close to the constrained edge,  with  several additional (secondary) stress concentration  regions within the panel. In the secondary  concentration regions in the panel, Case 1 has slightly   larger stress. Overall, the macroscipic response shows a  symmetry along the $Y$ direction, mainly due to the relative symmetric distribution of the pressure load and the geometry. 

While Case 1 has much higher temperature  than Case 2, the resulting plastic strain in the two simulations are of similar magnitude. This suggests that the spatial variation in the temperature field also promotes the accumulation of plastic strains. When  examining the  maximum dislocation discrepancy, $\Delta \rho _{\mathrm{tot, max}}$,  in the concentration region in the panel in Fig.~\ref{EffectofTemp}. It is observed that Case 2 has an $\Delta \rho _{\mathrm{tot, max}}$ value {5.7\%}  higher than that in Case 1. This suggests that the spatial variation in temperature  gradients  promotes the failure initiation  at the microscale, despite that common macroscopic quantities such as stress and plastic strains are largely similar. {While current experimental investigation of the panel response has not yet gone down to the subgrain level, it has been demonstrated that the presence of thermal gradients will shift the boundary layer transient, alter the panel response}~\citep{RileyZ:2017a} {and facilitate the plate buckling}~\citep{ThorntonE:1994a}. {The promotion of fatigue initiation by thermal gradient has also been widely observed for nickel-based alloy}~\citep{BrendelT:2008a, SunJ:2020a}. 

 {It should be noted that the current calibration is based on monotonic loading cases, while the cyclic response of the same material under pure cyclic loading has only been numerically studied}~\citep{LiuY:2020a}.  {Using the calibrated model from monotonic loading to a cyclic loading case, or outside the temperature and strain rates range may lead to potential deviation.  It would be beneficial to directly use cyclic stress-strain responses at a wide range of temperatures and strain rates to calibrate the model when such data become available, then the calibrated model could be widely used under different loading conditions.}

\subsection{Local response investigation and effects of texture}

In this section, we investigate the behavior in the areas of localized stress concentrations. In view of the anisotropy of the $\alpha$-dominated microstructure of the titanium alloy considered and   the strain partitioning due to non-uniform textures~\citep{EchlinM:2016a, MaR:2018a, LiuY:2021a},    we also study the effect of texture on the behavior of a panel subjected to dynamic loading. To introduce localizations, we add three bolt holes along edge \textit{AD}  of the panel as  shown in Fig.~\ref{PaneGeom} (b), and constrain the displacements of all nodes on the inner face of the bolt holes.     In addition to the original texture, we also investigate a case with a rotated texture. In the rotated texture case,   the third Euler angle of each   grain is subjected to reorient the dominant c-axis direction from along the panel length direction to along the width direction.  {We note that it is also possible to  have different regions of  the panel   to have different textures or  microstructures resulting from different manufacturing processes, such that high stress, strain and fatigue initiation locations can be altered and achieve different fatigue life}~\citep{LiuY:2020a}. 

\begin{figure}
	\centerline{
		\includegraphics[width=0.95\textwidth]{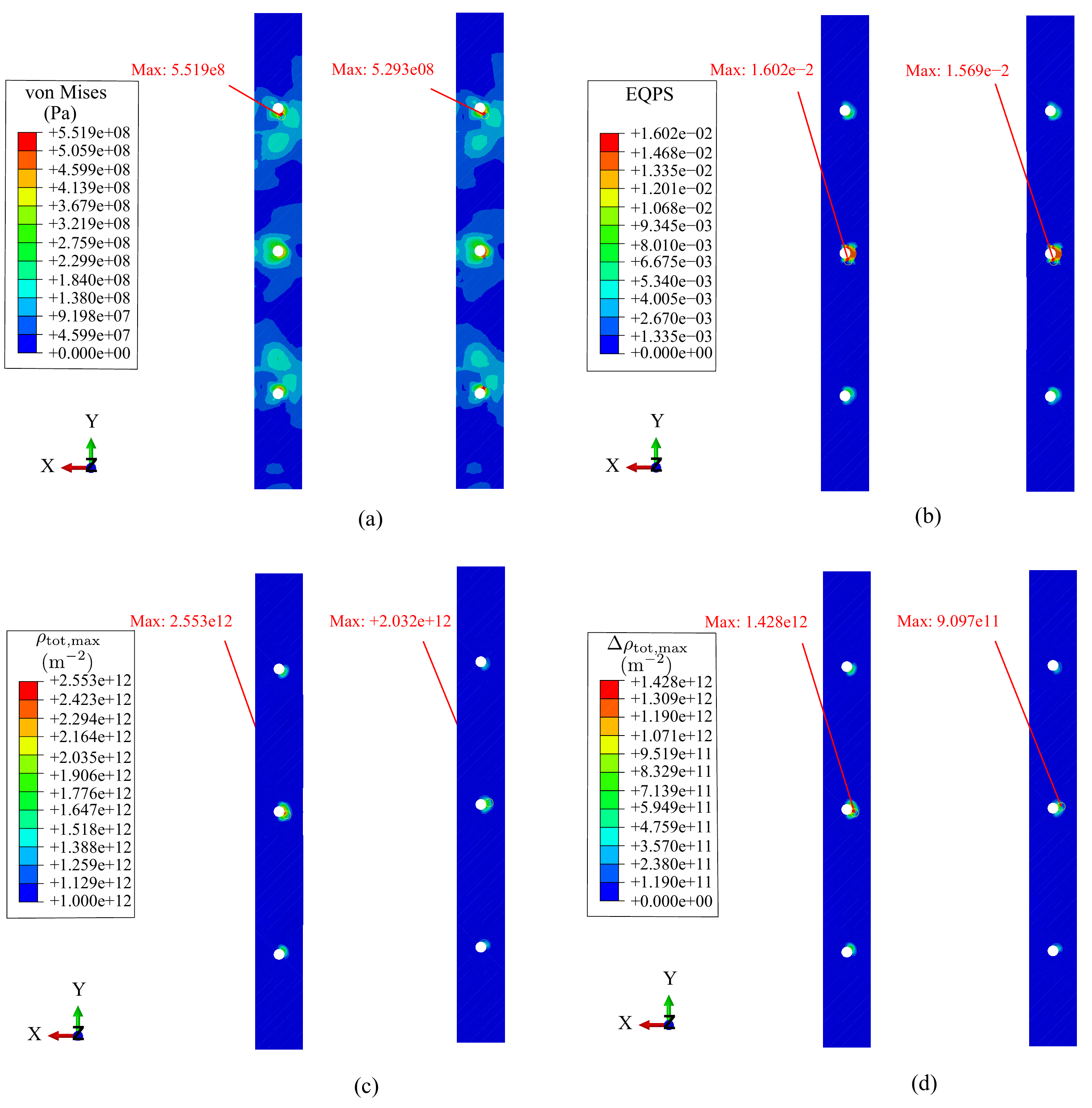}
	}
	\captionsetup{width=.8\textwidth}
	\caption{\small  Response of the bolted panel: a) von Mises stress, b) equivalent plastic strain and c)  maximum dislocation density discrepancy at the end of the simulation using the original (left) and rotated (right) texture.} 
	\label{BoltedPanelGlobal} 
\end{figure}

The von Mises, equivalent plastic strain and  maximum dislocation discrepancy around the holes  are shown in Fig.~\ref{BoltedPanelGlobal}. The stress concentrations are clearly seen along the three holes, and the plastic strain as well as the   maximum dislocation discrepancy all show higher values around the bolt holes.  The macroscale response measures such as  von Mises stress and $\bar \epsilon^{\mathrm{eqp}}$ show  very close agreement  between the two textures. However, for microscale response such as  $ \rho _{\mathrm{tot, max}}$ and $\Delta \rho _{\mathrm {tot, max}}$ show  significant differences.  This result demonstrates the presence of localized phenomenon, that are undetectable by homogenized measures. 

\begin{figure}[h]
	\centerline{
		\includegraphics[width=0.99\textwidth]{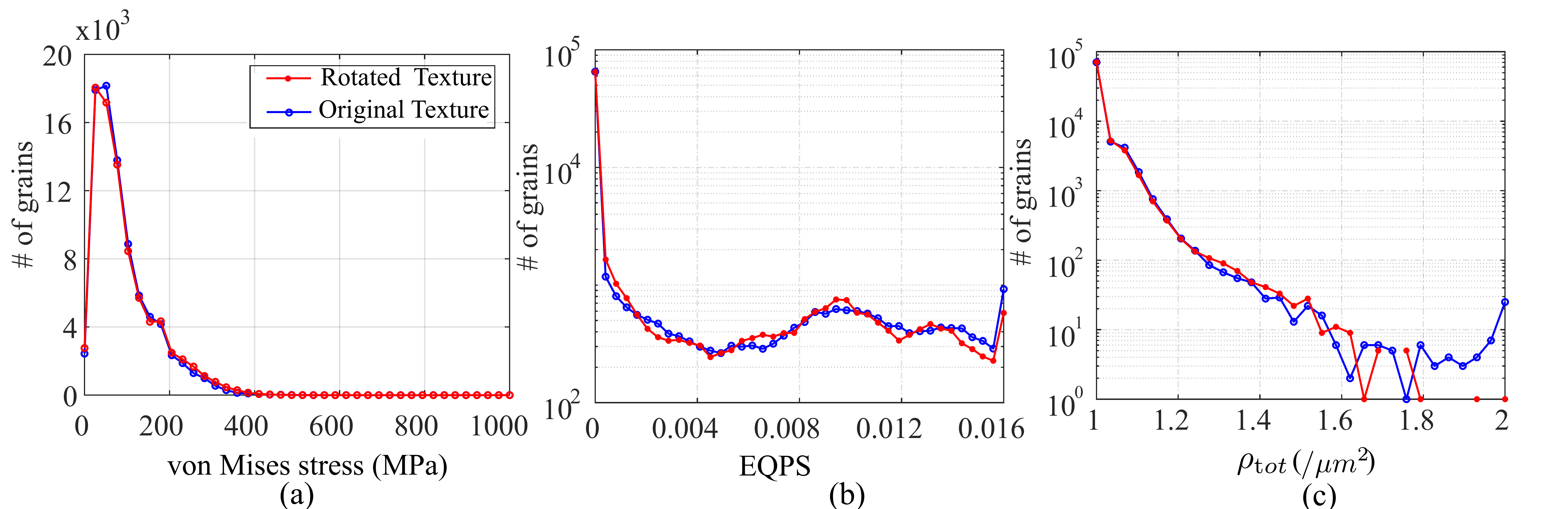}
 	}
	\captionsetup{width=.95\textwidth}
	\caption{\small  Local response of the region around the middle hole at the end of the simulation: a) von Mises stress, b) equivalent plastic strain and c)  dislocation density.  } 
	\label{BoltedPanelLocal} 
\end{figure}

\begin{figure}
	\centerline{
		\includegraphics[width=0.8\textwidth]{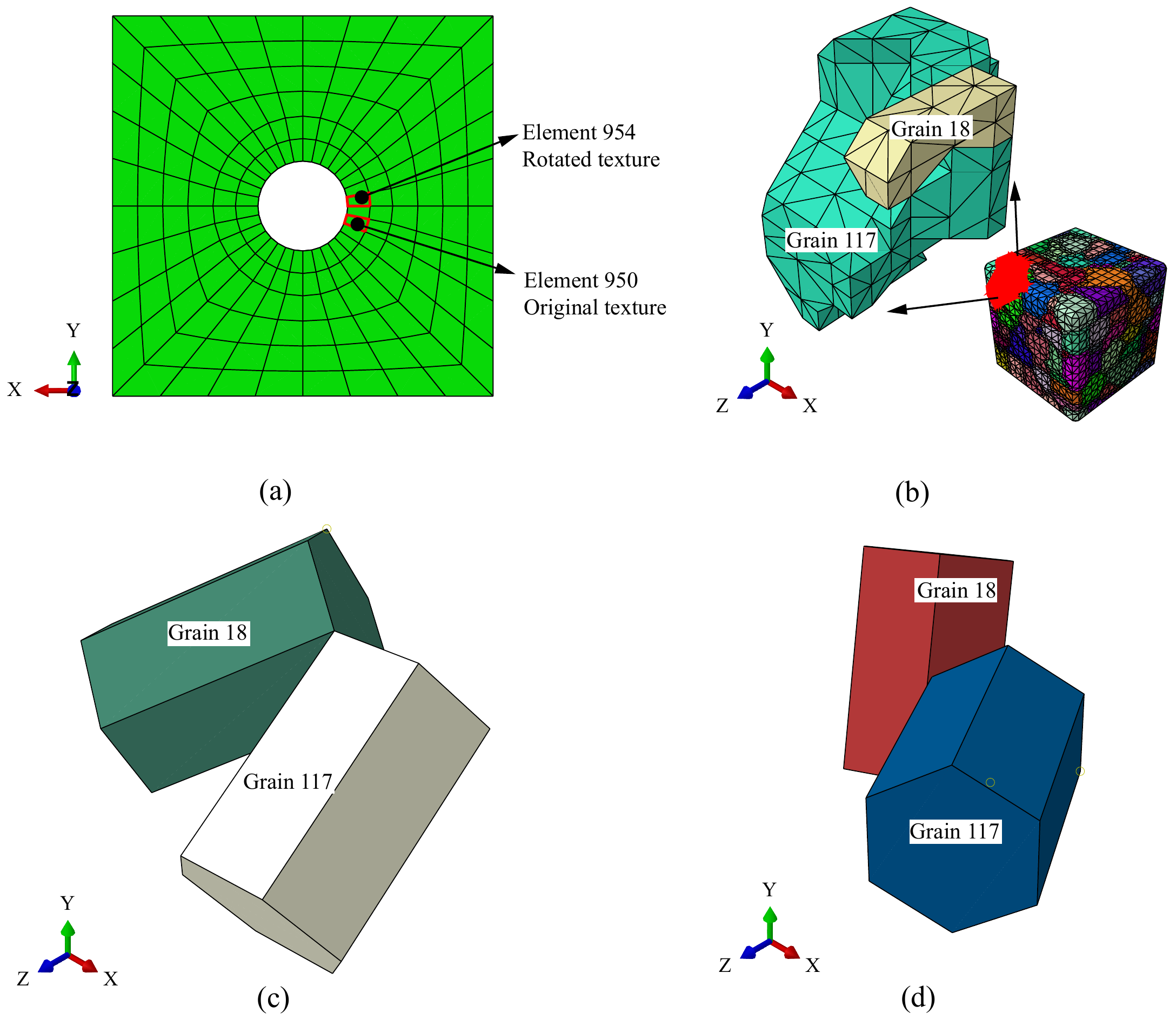}
	}
	\captionsetup{width=.8\textwidth}
	\caption{\small  Effects of the texture: a) locations of  $\Delta \rho_{\mathrm{tot, max}}$; b) Critical grain pair that has  $\Delta \rho_{\mathrm{tot, max}}$; c) Orientations of the critical grain pair for the original texture and; d)  Orientations of the critical grain pair for the rotated texture. } 
	\label{BoltedPanelLocalDiffTexture} 
\end{figure}

To better probe the local  response, we focus on a square region (2.54cm $\times$ 2.54 cm) around the center bolt hole (Fig.~\ref{BoltedPanelLocalDiffTexture} (a)).  This region contains 576 elements, hence   the response of 83,520 grains are tracked. The histograms of  stress, equivalent plastic strain and  dislocation density in each grain are plotted in Fig.~\ref{BoltedPanelLocal}.  Since the histograms have very long tails, the values beyond a certain magnitude are grouped into the last bin in the histogram for better visualization.

It is again clearly shown in  Fig.~\ref{BoltedPanelLocal} that while the von Mises stress and $\bar \epsilon^{\mathrm{eqp}}$ distributions are  close between the original and rotated texture, the $\rho_{\mathrm {tot}}$ have more significant differences, especially at the tail. For example, in the rotated texture case, there are only four grains that have   $\rho_{\mathrm {tot}}$ higher than 1.8 ${\mu m}^{-2}$. In contrast, there are more than twenty grains that above  $\rho_{\mathrm tot}$ higher than 2.0 ${\mu m}^{-2}$ in the original texture case. 


Figure~\ref{BoltedPanelLocalDiffTexture} (a) shows the locations in the panel, where the underlying microstructure has the   $\Delta \rho_{\mathrm{tot, max}}$ for the two textures. The locations  of the most critical points in the original and rotated texture cases are very close to each other,  and  close to the hole as shown in Fig.~\ref{BoltedPanelLocalDiffTexture}  (a).  The magnitude in the original texture case is approximately 50\% higher than that of the rotated texture. {It has been recognized both computationally and experimentally that texture has a significant role on the plastic response, and it is possible to use small clusters of similar crystal orientations, called micro-textured regions (MTR) or macrozones to change the fatigue sensitivities in titanium alloys}~\citep{EchlinM:2016a, LiuY:2021a}. {EHM could be potentially used to spatially design those MTR to improve the fatigue resistance of a given structural component.} We then further track down to identify the grain pairs that has the   maximum $\Delta \rho_{\mathrm {tot, max}}$. Interestingly, in both  cases, the dislocation discrepancy between the same two grains (grain 117 and 18) exhibits the $\Delta \rho_{\mathrm {tot, max}}$ as shown in 
Fig.~\ref{BoltedPanelLocalDiffTexture} (b). The orientations of this grain pair, are represented using  two 
hexagonal prisms as shown in Fig.~\ref{BoltedPanelLocalDiffTexture} (c) and (d), where the normal to the base of the prism is the c-axis of the grain. This example demonstrates the capability of EHM predicting failure initiation at a structural material point, by tracking the most critical grain pair inside the microstructure.    

\section{Conclusion}
This manuscript presented a multiscale reduced-order homogenization model for titanium structures  subjected to  thermo-mechanical loading associated with high-speed flight.  This development is an extension of the EHM model by incorporating thermal strain at the microscale, and accounts for temperature dependent material properties and evolution laws.   The developed model was fully calibrated and can accurately capture the stress-strain response of titanium materials under uniaxial tension across a wide range of temperature and strain rates.  A generic airplane skin panel subjected to thermo-mechanical loading associate with high-speed flight was then analyzed using the calibrated model to evaluate  to evaluate its response. The effects of temperature and microstructrue texture on the panel response were  also discussed. The following main conclusions are drawn from our investigations:
 \begin{itemize}
\item  The developed EHM model is fully capable of incorporating the thermal effects, and can accurately capture the stress-strain response of Ti6242S under a wide range of temperatures and strain rates. 
  \item The  developed EHM model can allow efficient microstructure-informed structural simulation, providing not only the macroscopic response, but also the response of the underlying microstructure associated with a macroscopic material point.  These localized response provides insights in localized response prediction, such as fatigue initiation.  

\item The simulation results from different temperatures  indicate  the spatial variation in
temperature field promotes the accumulation of plastic deformation, and dislocation density discrepancy
between neighboring grains, leading to earlier initiation of fatigue damage within the microstructure.

\item  The study on different texture indicates while the macroscopic response may only be slightly affected by the microstructure texture, the localized response highly depends on the microstructure and leads to very different fatigue initiation time.   Hence it is possible to delay the initiation of fatigue damage by proper design  of texture at critical locations.

 \end{itemize}

The developed EHM model for thermo-mechancial loading can be advanced in a number of fronts for  modeling of aerostructures subjected to high-speed flight environments. Firstly, the FIP used in this manuscript need to be compared with existing FIPs and further validated using experiments. This process will allow accurate fatigue initiation time and sites  prediction in large scale aerostructures.  Next, other physics processes associate with high-speed flight, such as oxygen ingress and oxygen-assisted material embrittlement  also plays significant role in determining structure  response. It is beneficial to incorporate these process in the EHM model. {In addition, we will also  focus  on a systematic study of the significance of the material and loading uncertainty associated with high-speed flight, and use of a  database of reduced-order models  for  Titanium with   selected microstructures   from different manufacturing processes to optimize the spatial variation of the microstructure for improved fatigue resistance under high-speed flight condition.} 


\section*{Funding Sources}
The authors gratefully acknowledge the research funding from the Air Force Office of Science
Research Multi-Scale Structural Mechanics and Prognosis Program (Grant No: FA9550-13-1-0104). We also acknowledge the technical cooperation of the Structural Science Center of the  Air Force Research Laboratory. X. Zhang has  been financially supported by the faculty start-up funding from the University of Wyoming, which is also gratefully acknowledged.

\section*{Appendix}
The coefficient tensors $\mathbf{A}$,  $\mathbf{P}$, and $\mathcal{A}^{(\beta)}$ are functions of elastic ($\mathbf{H}(T, \mathbf y)$),  phase inelastic ($\mathbf{h}^{(\alpha)}(T, \mathbf {y})$) and thermal ($\pmb{\mathcal{H}}(T, \mathbf{y})$) influence functions. At a given temperature $T$, the expressions for the coefficient tensors are
\begin{equation}
A^{(\beta)}_{ijkl}(T)=
I_{ijkl} +  \int_{\Theta^{(\beta)}}H_{(i,y_j), kl}(T, {\bf y})\,\mathrm{d}{\bf y} 
\label{ATen1}
\end{equation}
\begin{equation}
P^{(   \beta \alpha)}_{ijkl}(T)=\frac{1}{|\Theta^{(\beta)}|}\int_{\Theta^{(\beta)}} \int_{\Theta^{(\alpha)}}h_{(i, y_j)kl}(T, {\bf y, \hat{\bf y}) \,\mathrm{d}{\hat {\bf y}}} \,\mathrm{d}{\bf y}
\label{PartPTen}
\end{equation}
\begin{equation}
M^{( \alpha)}_{ijkl}(T)=\frac{1}{|\Theta^{\alpha}|}\int_{\Theta^{(\alpha)}}M_{ijkl}(T, {\bf y})\,\mathrm{d}{\bf y}
\label{MTen1}
\end{equation}
\begin{equation}
\mathcal{A}^{(\beta)}_{ij}( T)=\int_{\Theta}\mathcal{H}_{(i,y_j)}(T, {\bf y})\,\mathrm{d}{\bf y}
\label{TTen1}
\end{equation}
where $\mathbf{I}$ is the identity tensor, $\mathbf M$ the compliance tensor. Elastic and phase inelastic influence functions,  $\mathbf{H}(T, \mathbf {y})$, $\mathbf{h}(T, \mathbf y)$ and $\pmb{\mathcal{H}}(T, \mathbf{y})$ are calculated from the so called elastic influence function problem,   phase inelastic influence function problem and thermal influence function problem given in Box~\ref{BoxEIF}-\ref{BoxTIF}:

\renewcommand{\boxedtext}[3]{
	\begin{Box} \caption{\small{#1}}
		\hspace{1.cm}
		\begin{center}	
			\fbox{\begin{minipage}[c]{0.5\linewidth} 
					\small{#2}      
			\end{minipage}}
		\end{center}       
		\label{#3}
	\end{Box}
}
%

\vspace{-0.2cm}
\boxedtext{Elastic influence function problem.}{
	\begin{equation*}
	\begin{split}
	\{ L_{ijmn}(T, {\bf y})[H_{(m, y_n)kl}(T, {\bf y}) +I_{mnkl}] \}_{,y_j}=0; \qquad {\bf y} \in \Theta \\
	{\Theta} - \textrm{periodic boundary conditions on } \mathbf{y}\in \Gamma_{\Theta} \\
	\end{split}
	\label{EIF}
	\end{equation*}
}{BoxEIF}
%
\vspace{-0.5cm}
\renewcommand{\boxedtext}[3]{
	\begin{Box} \caption{\small{#1}}
		\hspace{1.cm}
		\begin{center}	
			\fbox{\begin{minipage}[c]{0.6\linewidth} 
					\small{#2}
			\end{minipage}}
		\end{center}       
		\label{#3}
	\end{Box}
}
%
\vspace{0.2cm}
\boxedtext{Phase inelastic influence function problem.}{
	\begin{equation*}
	\begin{split}
	\bigg\{L_{ijmn}(T, {\bf y})\Big[h_{(m, y_n)kl}(T, \mathbf{y}, \hat{\mathbf y})-I_{mnkl} \, \mathrm{d}({\mathbf y} - {\hat{\mathbf y}})\Big] \bigg\}_{,y_j}=0;  \hspace{0.3cm} \mathbf{y}, \hat{ \mathbf{y}} \in \Theta \\
	{\Theta} - \textrm{periodic boundary conditions on } \mathbf{y}\in \Gamma_{\Theta} \\
	\mathrm{d} -  \textrm{ the Dirac delta function}
	\end{split}
	\label{EIF}
	\end{equation*}
}{BoxPIF}


\vspace{-1.0cm}
\boxedtext{Thermal influence function problem.}{
	\begin{equation*}
	\begin{split}
 \Bigg\{L_{ijkl}(T, {\bf y}, T)  \bigg [ \mathcal{H}_{k, y_l}({\mathbf y}, T)-{ \alpha}_{kl}({\mathbf y})\bigg ]  \Bigg\}_{,y_j}
=0  \\
	{\Theta} - \textrm{periodic boundary conditions on } \mathbf{y}\in \Gamma_{\Theta} \\
	\end{split}
	\label{EIF}
	\end{equation*}
}{BoxTIF}



\end{document}